\RequirePackage{fix-cm}
\documentclass[smallextended]{svjour3}       
\pdfoutput=1
\newcommand*\GitHubLoc{https://www.qualikiz.com}
\newcommand*\GitHubLocSecond{https://gitlab.com/BartKremers/two-step-clustering}
\usepackage{graphicx}
\usepackage{amsmath}
\usepackage{amsfonts}
\usepackage{algorithm}
\usepackage{algorithmic}
\usepackage{xcolor}
\usepackage{booktabs}
\usepackage{array}
\usepackage{multirow}
\usepackage{tikz}
\usepackage{comment}
\usepackage{caption}
\usepackage{subcaption}
\usepackage[backend=biber,style=phys]{biblatex}
\usepackage{url,hyperref}
\DeclareGraphicsExtensions{.jpg,.png,.pdf}
\title{Two step clustering for data reduction combining DBSCAN and k-means clustering}
\subtitle{}

\author{B.J.J.~Kremers$^{1,a}$ \and
	A.~Ho$^{2}$      \and
	J.~Citrin$^{1,2}$  \and
	K.L.~van~der~Plassche$^{1,2}$
}

\institute{$^1$DIFFER, Eindhoven, the Netherlands \\
	$^2$Techincal University of Eindhoven, Eindhoven, the Netherlands \\
	$^a$\emph{Contact:} of B.J.J. Kremers \email{bjjkremers@gmail.com}           
}

\date{Received: date / Accepted: date}

\begin{document}

\maketitle

\begin{abstract}
A novel combination of two widely-used clustering algorithms is proposed here for the detection and reduction of high data density regions. The Density Based Spatial Clustering of Applications with Noise (DBSCAN) algorithm is used for the detection of high data density regions and the k-means algorithm for reduction. The proposed algorithm iterates while successively decrementing the DBSCAN search radius, allowing for an adaptive reduction factor based on the effective data density. The algorithm is demonstrated for a physics simulation application, where a surrogate model for fusion reactor plasma turbulence is generated with neural networks. A training dataset for the surrogate model is created with a quasilinear gyrokinetics code for turbulent transport calculations in fusion plasmas. The training set consists of model inputs derived from a repository of experimental measurements, meaning there is a potential risk of over-representing specific regions of this input parameter space. By applying the proposed reduction algorithm to this dataset, this study demonstrates that the training dataset can be reduced by a factor $\sim$20 using the proposed algorithm, without a noticeable loss in the surrogate model accuracy. This reduction provides a novel way of analyzing existing high-dimensional datasets for biases and consequently reducing them, which lowers the cost of re-populating that parameter space with higher quality data.
\end{abstract}

\section{Introduction}
\label{sec:Introduction}

Multiscale modelling is used to simulate complex phenomena in many areas of science and engineering~\cite{Groen2014}, such as material science~\cite{Suter2015}, bio-medicine~\cite{Tahir2014,Groen2013,Hoekstra2016}, engineering~\cite{Delgado-Buscalioni2003} and plasma physics. However, the evaluation of these multiscale models are generally computationally expensive. Within the field of plasma physics, recent developments have shown that targeted neural network (NN) surrogate models can reduce this computational time by orders of magnitude~\cite{VanDePlassche2020}. Unfortunately, acquiring a sufficiently representative dataset to train the NN in this application involves running the multiscale simulation itself. Thus, while NNs are fast to evaluate, they remain computationally expensive to develop~\cite{Alowayyed2019}. A more practical approach to building such a database with no \emph{a priori} information is to collect it from previous and current phenomenological studies that have been accumulated over time. This causes the data distribution to be determined by factors outside the considerations of surrogate model development and may result in the over-representation of certain regions in the parameter space, i.e. a biased dataset.


This paper proposes an algorithm for the detection and reduction of high data density space. In order to achieve this, a two-step procedure is proposed using the DBSCAN~\cite{Ester2009} and $k$-means~\cite{Stuart1982} algorithms. The DBSCAN algorithm is used to detect clusters with a user-defined minimum data density, as well as for initial outlier filtering. These clusters are then individually divided into $k$ evenly-dispersed groups by the $k$-means algorithm. The centroids of these groups are then used for data reduction by replacing each group with the actual data point closest to the identified centroid. The general difficulty in using the $k$-means algorithm alone is determining the value for $k$.

DBSCAN uses global parameters, defined in Chapter~\ref{sec:Algorithms}, which prevent the algorithm to differentiate clusters, or even regions within clusters, based on their densities. Several algorithms are proposed in literature to overcome these issues, which fall into two major categories: a partitioning-based or a non-partitioning-based approach~\cite{Kim2019}. A few partitioning-based algorithms include GMDBSCAN~\cite{Xiaoyun2008}, ADBSCAN~\cite{Khan2019}, PACA-DBSCAN~\cite{Jiang2011}, APSCAN~\cite{Chen2011}, AA-DBSCAN~\cite{Kim2019} and kAA-DBSCAN~\cite{Kim2019}. A few non-partitioning based approaches such as OPTICS~\cite{Ankerst1999} and DPC~\cite{Rodriguez2014}. However, DBSCAN also provides an estimate for the minimum data density of a cluster via its neighbourhood radius parameter, which can be subsequently used to advise a reasonable value for $k$.


To that effect, we propose a new approach for a flexible data reduction algorithm based on the detection of clusters with varying densities. The combination of algorithms proposed here is similar to a recently-proposed solution to the multiple circle detection problem~\cite{Scitovski2021}. However, the modifications made to the $k$-means algorithm to identify the centers of intersecting generalized circles does not serve the purpose of replacing a collection of data points with a reduced representative set. In addition, the iterative DBSCAN execution differs from currently available algorithms in that the clustering parameters are modified per iteration, such that it detects regions of increasing data density. As points are removed sequentially over each iteration, less internal memory is used for the detection and reduction of these regions. Due to this, the algorithm is more scalable to larger non-uniform datasets such as those compiled in a top-down fashion via data mining, as opposed to those built from the bottom-up using experimental design techniques.


Section~\ref{sec:Algorithms} provides an overview of the DBSCAN algorithm, the k-means algorithm and neural network surrogate modeling. In Section~\ref{sec:TwoStepAlgorithm} we propose the two step clustering algorithm for data reduction. In Section~\ref{sec:Application}, the algorithm is evaluated on two separate datasets: an artificially-generated test dataset for ease of visualization and an experimentally-derived dataset, generously provided by the Joint European Torus experimental fusion plasma device located in Culham, UK, to demonstrate an application of the algorithm in neural network surrogate modeling. While the input side of the datasets was selected, the corresponding output side for the supervised learning application was created using QuaLiKiz~\cite{Bourdelle2007}, a quasilinear gyrokinetics code for turbulence modeling in fusion plasmas. Section~\ref{sec:Conclusions} summarizes and concludes the paper.

\section{Properties of existing algorithms}
\label{sec:Algorithms}


These descriptions are based on a data-set, $D$, in a $d$-dimensional space, $\mathbb{R}^d$, consisting of $n$ objects.

\subsection{DBSCAN}
\label{subsec:DBSCAN}

DBSCAN is an algorithm for density-based clustering originally designed to discover clusters, $C$, of arbitrary shapes in spatial data~\cite{Ester2009}. The algorithm takes two input parameters: \begin{itemize}
	\item $\epsilon$, the radius of a hypersphere drawn around each point, known as a neighborhood, $N_{\epsilon}$;
	\item $MinPts$, the minimum number of points in the neighbourhood needed to be defined as a part of a cluster, including the current point.
\end{itemize}
With this spatial definition, the relations between the points, $p,q \in D$, can be categorized as follows:

\begin{definition}{(Directly density-reachable)}
	Object $p$ is considered directly density-reachable from object $q$ if $p \in N_{\epsilon}(q)$.
\end{definition}
\begin{definition}{(Density-reachable)}
	Object $p$ is considered density-reachable from object $q$, where $p,\,q \in D$, if a chain of objects, $\{p_1,\dots,p_{k+1}\} \in D$, exists such that $p_{i+1}$ is directly density-reachable from $p_i$, where $i \in [1,k]$, $p_1 = p$ and $p_{k+1} = q$.
\end{definition}
\begin{definition}{(Cluster, $C$)}
	A subset of $D$ where $|C| \geq MinPts$ and all objects $p \in C$ are density-reachable from all other objects $q \in C$.
\end{definition}

DBSCAN then finds these clusters by labelling points using the following definitions~\cite{Ester2009}:

\begin{definition}{(Core object, $x_{\text{core}}$)}
	If object $x$ is such that $|N_{\epsilon}(x)| \geq MinPts$, then $x$ is considered a core object.
\end{definition}
\begin{definition}{(Border object, $x_{\text{border}}$)}
	If object $x$ is directly density-reachable from a core object but $|N_{\epsilon}(x)| < MinPts$, then $x$ is considered a border object.
\end{definition}
\begin{definition}{(Noise object, $x_{\text{noise}}$)}
	If object $x$ is neither a core nor a border object, then $x$ is considered a noise object.
\end{definition}
Once the entire dataset is labelled, it groups together all points.

\subsection{$k$-means}
\label{subsec:kmeans}

The $k$-means algorithm is designed to partition a dataset into regions or clusters, $R$. This algorithm accepts a single input parameter:
\begin{itemize}
	\item $k$, the number of clusters into which the data is partitioned.
\end{itemize}
Once the value is set, $k$ centroids, $\{m_1,\dots,m_k\}$, are determined and each object, $x$, is associated to the nearest centroid. This determination is done by minimizing the spatial variance between the centroids and their associated data points.

Initially, the centroids are randomly chosen and the algorithm iterates two processes over a counter, $i$, until the centroids no longer change. In the first process, each object is assigned to the cluster, $R^{(t)}$, with the least squared Euclidean distance away from it. This can be mathematically expressed as follows:
\begin{equation}
	R_i^{(t)} = \{x : ||x - m_i^{(t)}||^2 \leq ||x - m_j^{(t)}||^2 \; \forall \; j \in [1,k] \}
\end{equation}
where each point is assigned to exactly one cluster. In the second process, the cluster centroids are recalculated based on the mean of the objects assigned to it. This can be expressed as:
\begin{equation}
	m_{i+1}^{(t)}=\frac{1}{|R_i^{(t)}|}\sum_{R_i^{(t)}} x
\end{equation}

\subsection{Neural networks}
\label{subsec:NeuralNetwork}

A neural network (NN) is a universal approximator consisting of nodes, each having a weight $w_j$ and a bias $b_j$. By interconnecting layers of these nodes and designing the connections purposefully, the NN can be made suitable for both regression or classification tasks. These weights and biases have to be learned by optimizing an objective function, $\mathcal{U}$. Typically, this objective function minimally consists of a goodness-of-fit term to fit the training dataset and a regularization term to avoid overfitting.

This study focuses on a previous fully-connected feed-forward NNs (FFNN), used to reproduce the input-output mapping of the quasilinear gyrokinetic transport model QuaLiKiz~\cite{Citrin2017,Bourdelle2016} for turbulence predictions in fusion plasmas. A custom architecture and cost function was used for training these NNs~\cite{VanDePlassche2020}, which were designed to better capture the desired physical characteristics of the problem. 

\section{Two step clustering algorithm}
\label{sec:TwoStepAlgorithm}

\begin{figure}[tb]
	\centering
	\begin{subfigure}[t]{0.45\textwidth}
		\centering
		\includegraphics[scale=0.35,trim={4cm 8cm 3cm 8cm},clip]{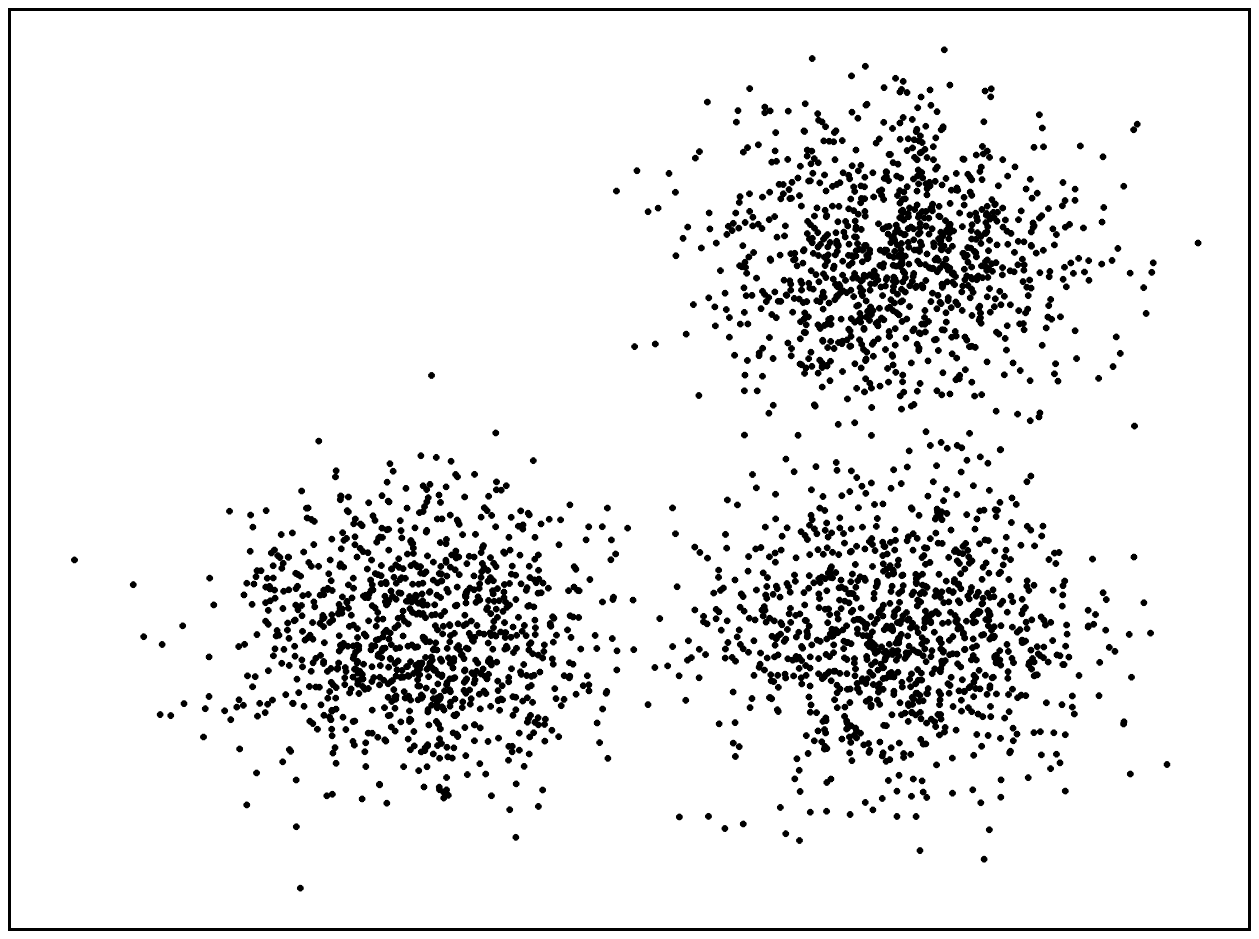}  
		\caption{Original dataset with three dense clusters and noise.}
		\label{fig:TwoStepSingle:Original}
	\end{subfigure}%
	~~~
	\begin{subfigure}[t]{0.45\textwidth}
		\centering
		\includegraphics[scale=0.35,trim={4cm 8cm 3cm 8cm},clip]{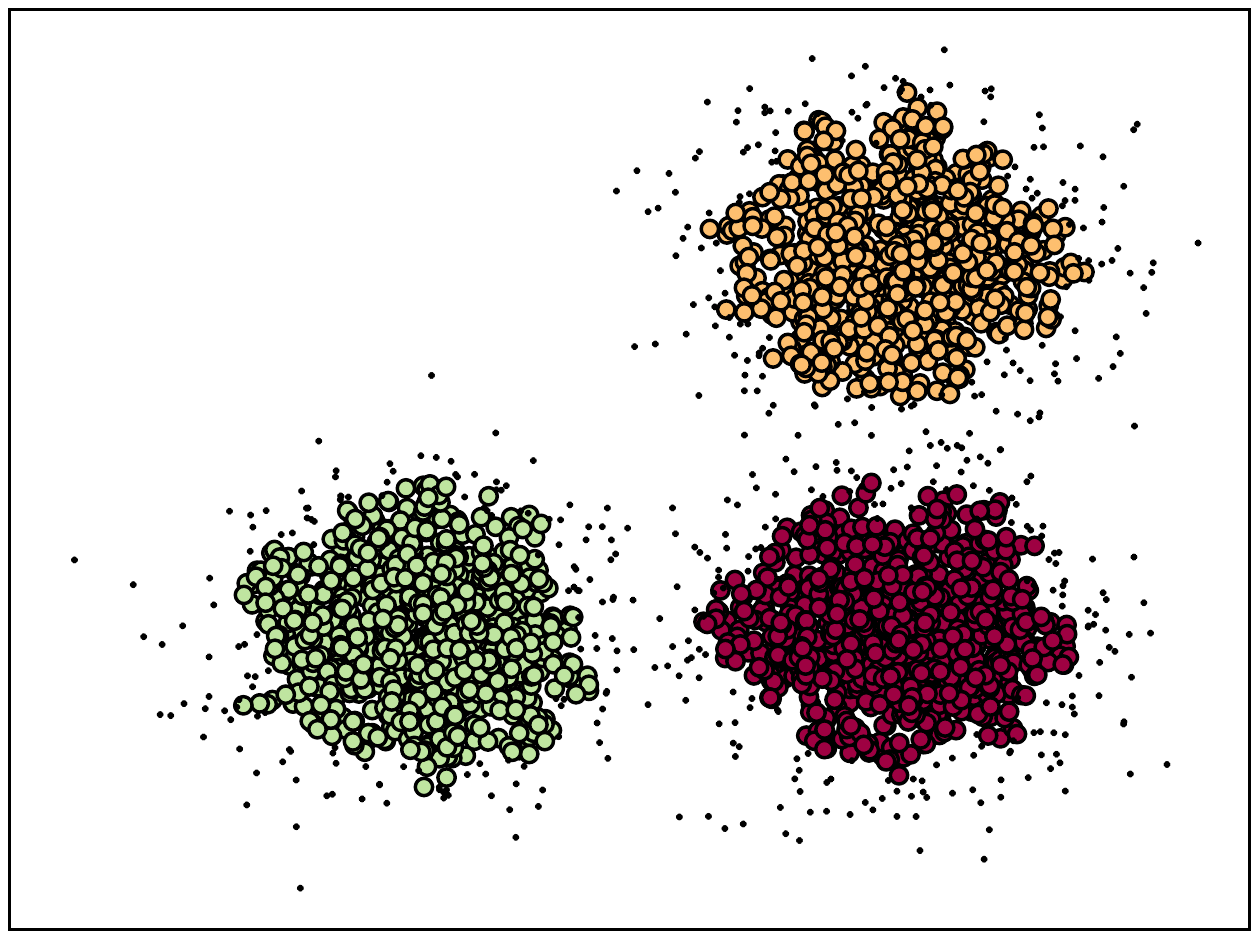}  
		\caption{Three clusters depicted in yellow green and red are detected by the DBSCAN algorithm. Black points are considered noise that do not fall into any cluster.}
		\label{fig:TwoStepSingle:DBSCAN}
	\end{subfigure}

	\begin{subfigure}[t]{0.45\textwidth}
		\centering
		\includegraphics[scale=0.35,trim={4cm 8cm 3cm 8cm},clip]{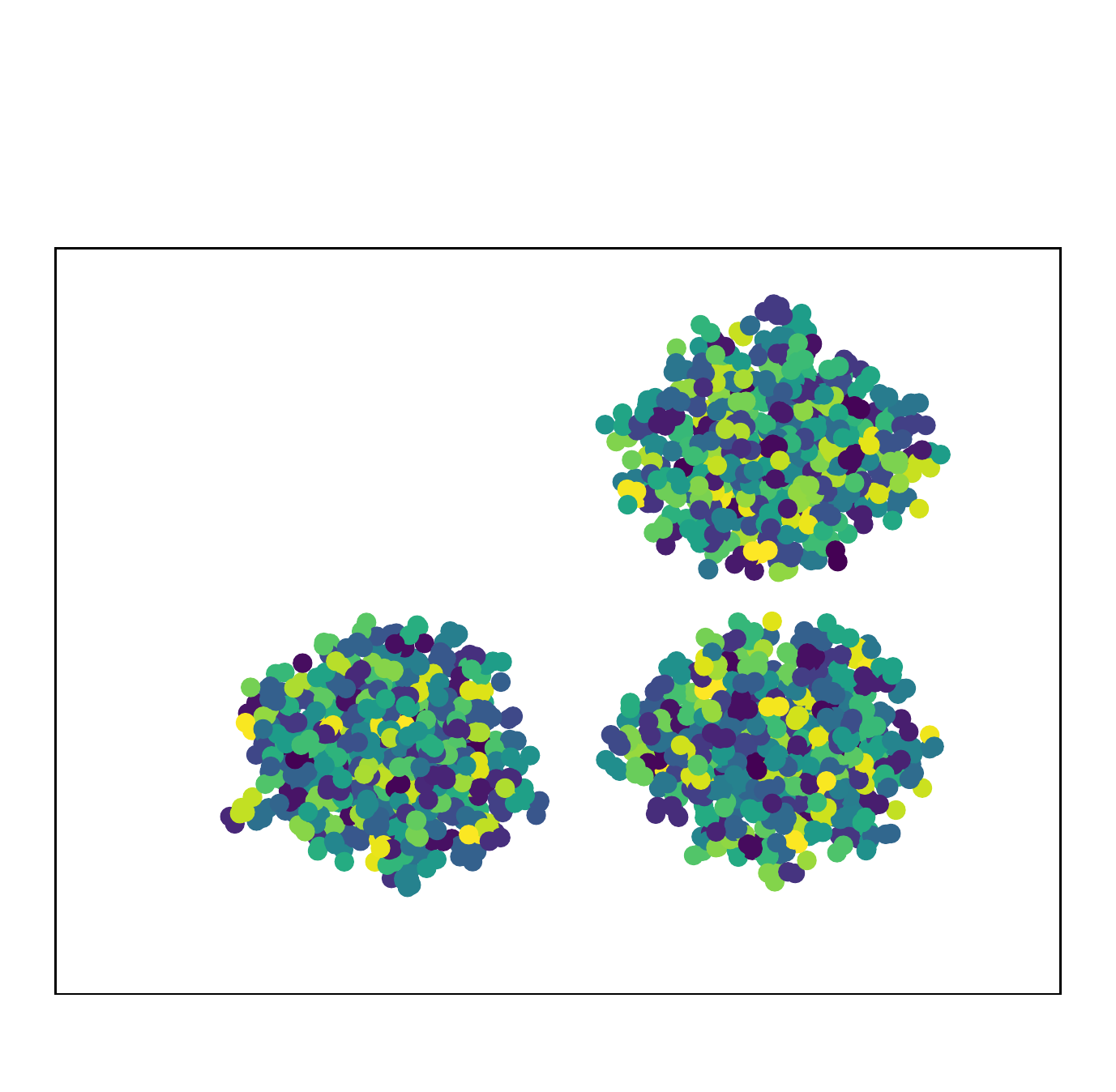}  
		\caption{Using $k$-means on the individual DBSCAN clusters the clusters are split into smaller partitions. Noise points are removed from the data-set.}
		\label{fig:TwoStepSingle:kmeans}
	\end{subfigure}%
	~~~
	\begin{subfigure}[t]{0.45\textwidth}
		\centering
		\includegraphics[scale=0.35,trim={4cm 8cm 3cm 8cm},clip]{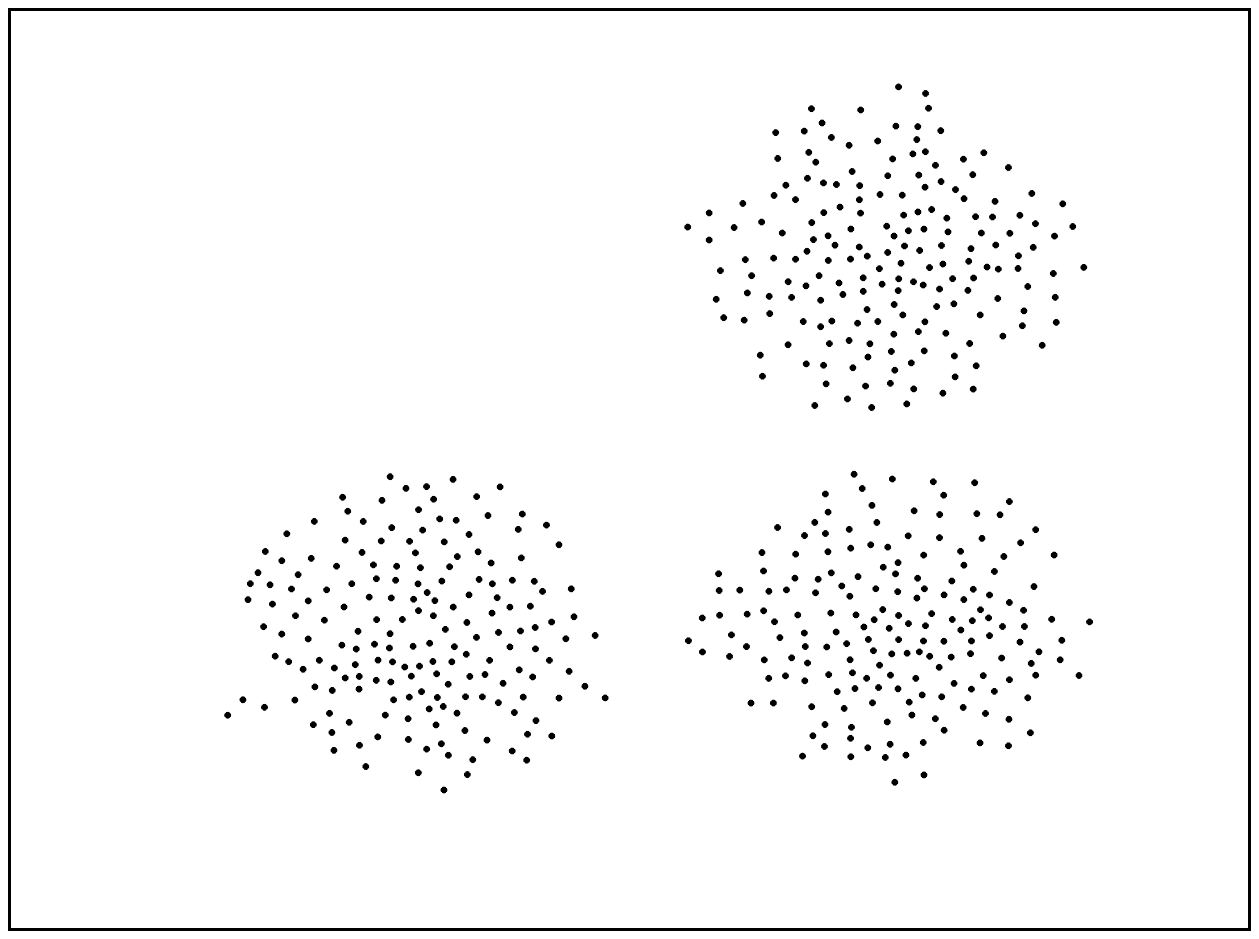}  
		\caption{Final data-set, k-means clusters replaced by point closest to the mean of the cluster.}
		\label{fig:TwoStepSingle:Reduced}
	\end{subfigure}
	\caption{Schematic example of two-step algorithm, with data cluster detection using DBSCAN and cluster reduction using k-means.}
	\label{fig:TwoStepSingle}
\end{figure}

This section describes the proposed two-step algorithm, firstly by explaining the concept behind combining the DBSCAN and $k$-means algorithms and then expanding its application into an iterative procedure. Figure~\ref{fig:TwoStepSingle} depicts the basic steps of the algorithm on a toy example, containing 3 densely populated regions in a 2D space. While not explicitly shown in the figure, it is often convenient to scale and translate the data ranges such that they have unit variance and zero mean. Starting with a dataset, $D$, as shown in Figure~\ref{fig:TwoStepSingle:Original}, the general two-step clustering algorithm goes through the following steps:
\begin{enumerate}
  \item Apply DBSCAN algorithm to a dataset, $D$, to determine regions, $C$, of high data density, resulting in Figure \ref{fig:TwoStepSingle:DBSCAN}.
  \item Label all points where $D \notin C$ as $C_{noise}$ and discard.
  \item Apply $k$-means algorithm to each individual cluster, $C$, splitting it into $k$ regions, $R_C$, resulting in Figure \ref{fig:TwoStepSingle:kmeans}.
  \item The nearest data points, $x \in D$, to the $k$-means cluster centroids are determined. These points become the output reduced dataset, $D'$, as shown in Figure \ref{fig:TwoStepSingle:Reduced}.
\end{enumerate}
A summary of the input parameters to the algorithm can be found in Table~\ref{tbl:RecommendedInputParameterValues}.

While these steps form the basis for the two-step clustering algorithm, the application also aims to distinguish between regions of varying local data density, $d$. Thus, this two-step algorithm is iterated over $i$ steps while decreasing the radius, $\epsilon_i$, in order to find increasingly dense regions in $D_i$. In order to improve the robustness of the final reduced dataset, the first iteration of this loop doubles as a noise filter. For this reason, it was necessary to decouple the neighbourhood radius used in the initial pass, $\epsilon_0$, and the neighbourhood radius used in the first real iteration, $\epsilon_1$.

For the purposes of this algorithm, the local density was defined as:
\begin{equation}
\label{eq:DensityExpression}
	d = \frac{\bar{n}}{V\!\left(\epsilon_i, N\right)}
\end{equation}
where $\bar{n}$ is the average number of neighbours over all $x_{core} \in C$, $V$ is the volume of an $N$-dimensional ball, and $N$ is the dimensionality of the dataset $D$. This allows the $k$-means parameter, $k$, to be chosen adaptively for each density bin. The adaptive selection of $k$ was implemented using the following expression:
\begin{equation}
\label{eq:NumberOfClustersRealExpression}
	\tilde{k} = a n \left( 1 - \zeta \cdot \left[ 1 - e^{-\tau d} \right] \right) \, , \quad \tau \in \left(0, \infty\right)
\end{equation}
\begin{equation}
	\label{eq:NumberOfClustersIntegerExpression}
	k = \begin{cases}
		\left\lfloor \tilde{k} \right\rfloor, & \text{for } \tilde{k} > 1 \\
		1, & \text{otherwise}
	\end{cases}
\end{equation}
where $a$ is a fractional value representing the desired global reduction ratio, $\zeta$ is a weighting factor between a uniform reduction scheme and a density-dependent reduction scheme, and $\tau$ is a control parameter setting how quickly $k \to 1$ in the density-dependent scheme.

The form of Equation~\eqref{eq:NumberOfClustersRealExpression} was chosen purely on functionality. By examining the limits of $\zeta$, it can be easily seen that choosing $\zeta = 0$ enforces $k = an$ for all clusters. This selects a uniform reduction ratio throughout the entire final dataset based on the user-defined value of $a$. Choosing $\zeta = 1$ causes $k$ to be determined solely by the last term in Equation~\eqref{eq:NumberOfClustersRealExpression}. The combination of Equations~\eqref{eq:NumberOfClustersRealExpression} and \eqref{eq:NumberOfClustersIntegerExpression} can be dissected further by examining the limits of $\tau$. As $\tau \to 0$, $k = an$ and a uniform reduction ratio is again applied. However, as $\tau \to \infty$, $\tilde{k} \to 0$ and $k \to 1$ meaning that all DBSCAN clusters are reduced to a single centroid. In general, it is recommend to have a reasonably small but finite value for $\tau$, such that $k \to 1$ gradually as $d$ increases, meaning a higher reduction ratio for more dense clusters. 

In addition to the three parameters described in Equation~\eqref{eq:NumberOfClustersRealExpression} replacing $k$ and $\epsilon_0$ replacing $\epsilon$, another six input parameters are introduced to the algorithm:
\begin{itemize}
	\item $a$, the global reduction ratio, which also defines the minimum reduction factor;
	\item $a_{noise}$, the reduction ratio to be applied to $\left\lbrace x_{noise} \right\rbrace$ in all iterations except the initial denoising pass;
	\item $\epsilon_1$, the starting neighbourhood radius to be used after the initial denoising pass;
	\item $\Delta\epsilon$, the step size applied to $\epsilon_1$ between consecutive iterations after the initial denoising pass;
	\item $n_\text{min}$, the number of points in a cluster below which the $k$-means reduction step is not applied to the cluster;
	\item $n_\text{max}$, the number of points in a cluster above which it is considered too large to be reduced in a given iteration.
\end{itemize}

The following definitions are used for the reduction and iterations:
\begin{definition}{(Saved cluster $C_s$)}
	Clusters $C$ where $MinPts \leq |C| \leq n_\text{min}$ are defined as saved clusters and transferred directly into the final dataset, $D_f$. It is assumed to contain sufficiently valuable information due to its sparseness.
\end{definition}
\begin{definition}{(Reduced cluster $C_r$)}
	Clusters $C$ where $n_\text{min}< |C| \leq n_\text{max}$ are defined as reduced clusters, to which the $k$-means reduction step is applied. The data points in these clusters nearest to the $k$-means centroids, $D'$, are transferred to the final dataset, $D_f$.
\end{definition}
\begin{definition}{(Passed cluster $C_p$)}
	Clusters $C$ where $n_\text{max} < |C|$ are defined as passed clusters as they contain sufficiently high data density to be subject to additional iterations with lower $\epsilon$. None of the points in these clusters are transferred to the final dataset, $D_f$, in the current iteration.
\end{definition}

After the initial DBSCAN execution, the set of noise points, $\left\lbrace x_{noise} \right\rbrace$ are discarded, $C_s$ are added to the final database, and the k-means algorithm is applied to reduce the individual reduction clusters, $C_r$. The $C_p$ are then passed to the next iteration DBSCAN, where the procedure is slightly modified compared to the initial pass. Specifically, $\left\lbrace x_{noise} \right\rbrace$ are now passed as an independent cluster to the k-means algorithm using a $k$ determined by $a_{noise}$ and $\zeta=0$. The resulting reduced set of noise points are added to the $C_s$ and are thus added to the final database $D_f$. Each following iteration reduces $\epsilon$ by $\Delta\epsilon$ meaning DBSCAN selects out regions of increasingly higher data density from the dataset. The iterations continue until either all of the original data has passed through the $k$-means reduction process or a user-defined maximum density is reached. This maximum density is determined through a minimum neighbourhood radius, $\epsilon_\text{min}$, and was implemented to prevent extremely dense regions from slowing down the algorithm and biasing the final data-set, $D_f$. 

\begin{algorithm}[H]
	\caption{Two-step clustering procedure}
	\begin{algorithmic}
		\STATE Load and rescale $D$
		\STATE $D_i=D$
		\WHILE{$D_i$ not empty}
			\STATE DBSCAN $D_i$
			\STATE Define $D_n = \{x_{noise}\}$
			\IF{initial pass}
				\STATE Remove $D_n$ from $D_i$
			\ELSE
				\STATE Calculate adaptive $k$ with $a = a_{noise}$, $\zeta=0$
				\STATE Apply $k$-means to $D_n$
				\STATE Find $D' = \{x_m : \min |x_m - m_i| \; \forall \; i \in \left[1,k\right]\}$
				\STATE Move $D'$ to $D_f$
				\STATE Remove remaining $x \in D_n$ from $D_i$
			\ENDIF
			\FOR{$C_s$}
				\STATE Move $C_s$ to $D_f$
			\ENDFOR
			\FOR{$C_r$}
				\STATE Calculate $d$ and adaptive $k$
				\STATE Apply $k$-means to $C_r$
				\STATE Find $D' = \{x_m : \min |x_m - m_i| \; \forall \; i \in \left[1,k\right]\}$
				\STATE Move $D'$ to $D_f$
				\STATE Remove remaining $x \in C_r$ from $D_i$
			\ENDFOR
			\STATE Decrement $\epsilon$ by $\Delta\epsilon$
			\IF{$\epsilon \le \epsilon_\text{min}$}
				\STATE Calculate $d$ and adaptive $k$
				\STATE Apply $k$-means to $D_i$
				\STATE Find $D' = \{x_m : \min |x_m - m_i| \; \forall \; i \in \left[1,k\right]\}$
				\STATE Move $D'$ to $D_f$
				\STATE Remove all remaining $x \in D_i$ from $D_i$
			\ENDIF
		\ENDWHILE
		\STATE Save $D_f$
	\end{algorithmic}
	\label{alg:TwoStepAlgorithm}
\end{algorithm}

\begin{table}[tb]
	\centering
\begin{tabular}{|l|l|l|l|}
	\hline
	\textbf{Parameter} & \textbf{Symbol} & \textbf{Range} & \textbf{Recommended} \\\hline
	Initial pass neighborhood radius & $\epsilon_0$ & $\mathbb{R} \in \left(0, \infty\right)$ & 1.0 \\\hline
	First iteration neighborhood radius & $\epsilon_1$ & $\mathbb{R} \in \left(0, \infty\right)$ & 0.75 \\\hline
	Neighbourhood radius step size & $\delta\epsilon$ & $\mathbb{R} \in \left(0, \infty\right)$ & 0.05 \\\hline
	Minimum neighborhood radius & $\epsilon_\text{min}$ & $\mathbb{R} \in \left(0, \infty\right)$ & 0.01 \\\hline
	Minimum cluster size & $MinPts$ & $\mathbb{Z} \in \left[2, \infty\right)$ & $\le 10$ \\\hline
	Minimum reduced cluster size & $n_\text{min}$ & $\mathbb{Z} \in \left[2, \infty\right)$ & same as $MinPts$ \\\hline
	Maximum reduced cluster size & $n_\text{max}$ & $\mathbb{Z} \in \left[2, \infty\right)$ & 100 \\\hline
	Reduction scheme parameter & $\zeta$ & $\mathbb{R} \in \left[0, 1\right]$ & 1.0 \\\hline
	Density-dependent factor & $\tau$ & $\mathbb{R} \in \left(0, \infty\right)$ & 0.001 \\\hline
\end{tabular}
	\caption{Input parameters for the two-step reduction algorithm and their recommended starting values.}
	\label{tbl:RecommendedInputParameterValues}
\end{table}

At this point, it is important to note that the recommended value for $n_\text{min}$ is the same value as $MinPts$. This effectively removes the presence of saved clusters, $C_s$, within the algorithm. In practice, the presence of these clusters tended to significantly hinder the efficacy of the density-dependent reduction scheme, meaning the $\zeta=0$ behaviour closely resembled that of $\zeta=1$. However, this functionality is retained in the code in the case that another application has need for it.

\section{Demonstrative Applications}
\label{sec:Application}

In order to demonstrate the capability of the two-step clustering algorithm described in Section~\ref{sec:TwoStepAlgorithm}, it was applied to two test cases. The first is a 3D sample dataset created from 2 input parameters and 1 output parameter from the QuaLiKiz turbulent transport code, described in Section~\ref{subsec:TestCase3D}. Although the full input-output dimensionality of the code is much higher, this selection allows for an easy visualization of the results from the clustering algorithm. The second test case is of a 16D dataset, 15 inputs and 1 output, described in Section~\ref{subsec:TestCase16D}. This dataset is based on the same experimental data used in the previous NN development study using this physics code~\cite{Ho2021}. Additional NNs were trained using the final reduced dataset, $D_f$, from the two-step algorithm and compared to the original NN.

At this point, it should be noted that the specific choice to include an output parameter is these test cases is due to the application on NN regression training datasets. The algorithm itself does not functionally distinguish between the various columns in the provided data table.

\subsection{3D test dataset}
\label{subsec:TestCase3D}

The 3D dataset consists of two input parameters, namely the logarithmic electron temperature gradient, $R/L_{Te}$, and the logarithmic ion temperature gradient, $R/L_{Ti}$, and one output parameter, the ion temperature gradient (ITG) ion heat flux, $q_{i,\text{ITG}}$. In order to test the density-dependent functionality of the proposed two-step clustering algorithm, the input half of the dataset was sampled explicitly to contain a variable data density. The input parameters were sampled such that $R/L_{Te}, R/L_{Ti} \in \left[0,18\right]$, using a base uniform distribution with Gaussians added on top. The locations for these Gaussian distributions are $(4.5,4.5)$, $(13.5,4.5)$, $(4.5,13.5)$, $(13.5,13.5)$ with an increasing maximum probability per addition. These sampled inputs are then inserted into QuaLiKiz to calculate $q_{i,\text{ITG}}$, with the other settings provided in Appendix. The resulting 3D dataset is shown in Figure~\ref{fig:3DOriginal}.

\begin{figure}[tb]
	\centering
	\includegraphics[scale=0.7]{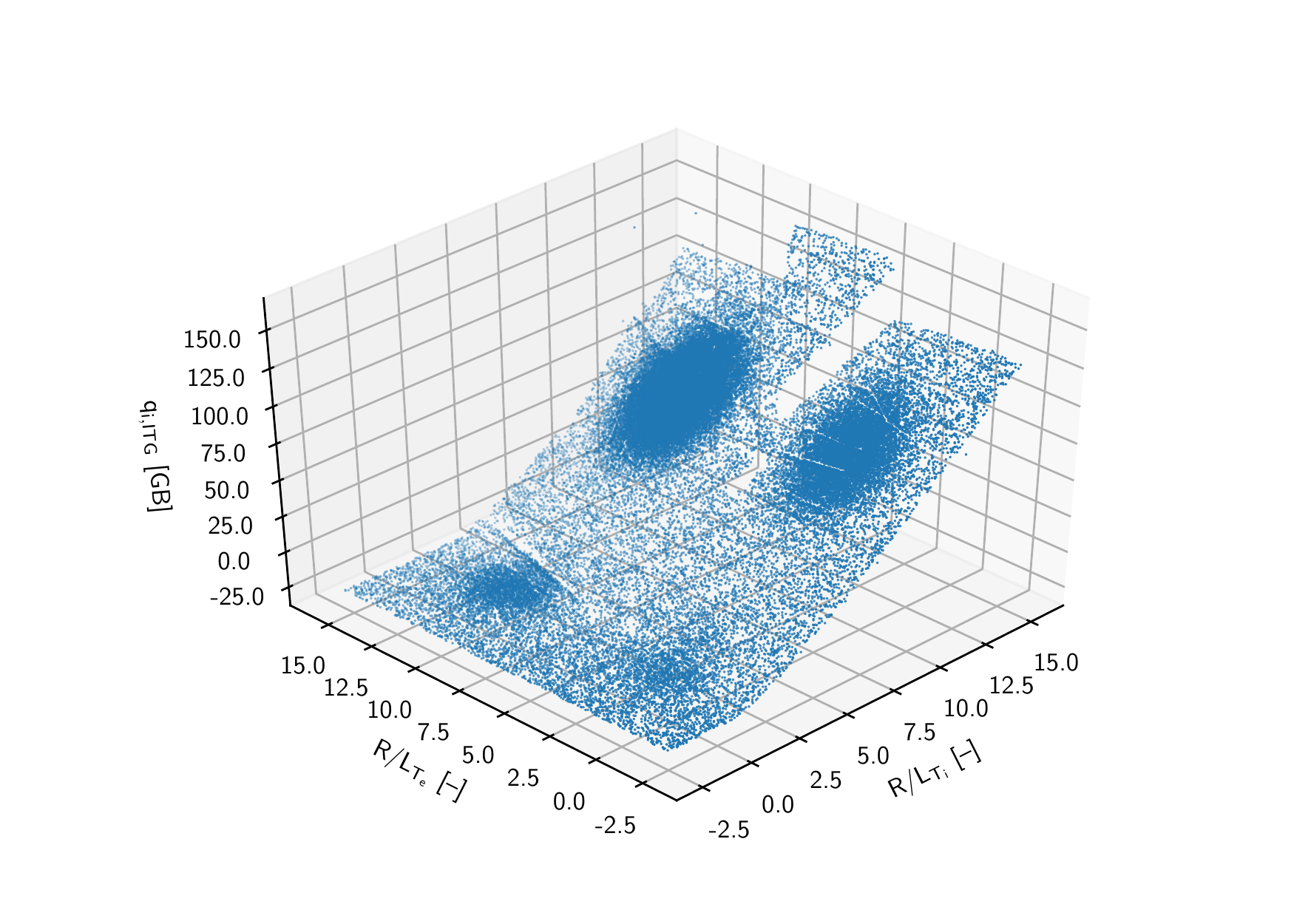}
	\caption{Test 3D dataset (two-input, one-output) generated via random sampling on the inputs, $R/L_{Te}$ and $R/L_{Ti}$, and using QuaLiKiz to compute the output, $q_{i,\text{ITG}}$, with normalized units (GB). The input space was sampled within (0,18) using a fixed uniform density with four Gaussian peaks with varying peak data densities added on top.}
	\label{fig:3DOriginal}
\end{figure}

In principle, the algorithm is capable of including multiple output dimensions but this demonstration only uses one for ease of interpretation.

\subsubsection{3D dataset reduction}
\label{subsubsec:Reduction3D}

The two-step clustering algorithm is then applied to the 3D dataset described in Section~\ref{subsec:TestCase3D}. A scan of the various input parameters of the algorithm results in different overall reduction factors, as expected, where the overall reduction factor, $f$, is defined in this study as follows:
\begin{equation}
\label{eq:ReductionFactor}
	f = \frac{n_{\textrm{orig}}}{n_{\text{red}}}
\end{equation}
where $n_{\textrm{orig}}$ is the number of points in the original dataset and $n_{\text{red}}$ is the number of points in the reduced dataset.

\begin{figure}[tb]
	\centering
	\includegraphics[scale=0.7]{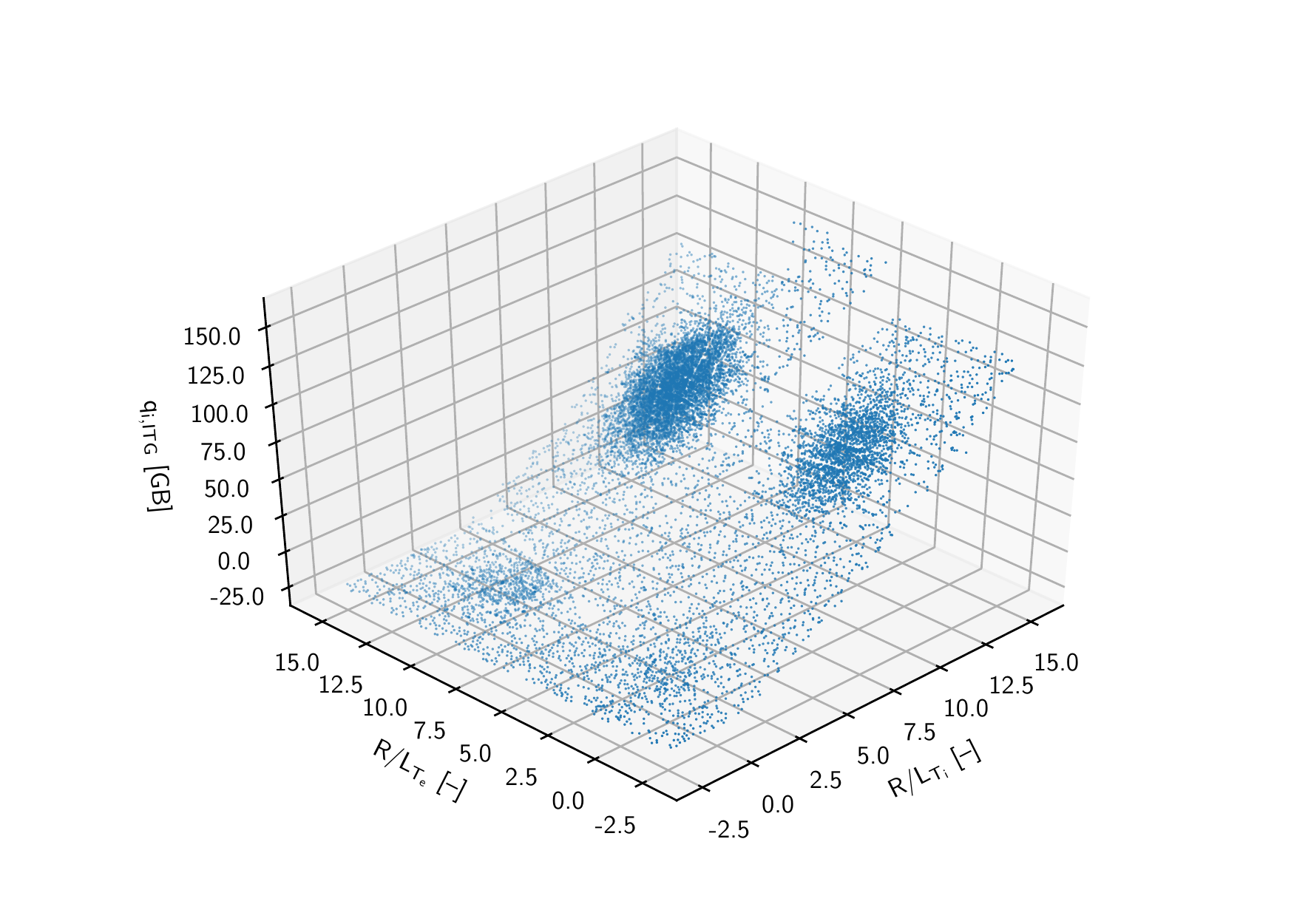}
	\caption{Reduced 3D sample dataset using $\zeta=0$, applying an overall reduction factor, $f=5.2$. As shown, the density distributions in the original dataset are mostly kept and all four high-density regions visually distinguishable.}
	\label{fig:3DReductionUniform}
\end{figure}

\begin{figure}[tb]
	\centering
	\includegraphics[scale=0.7]{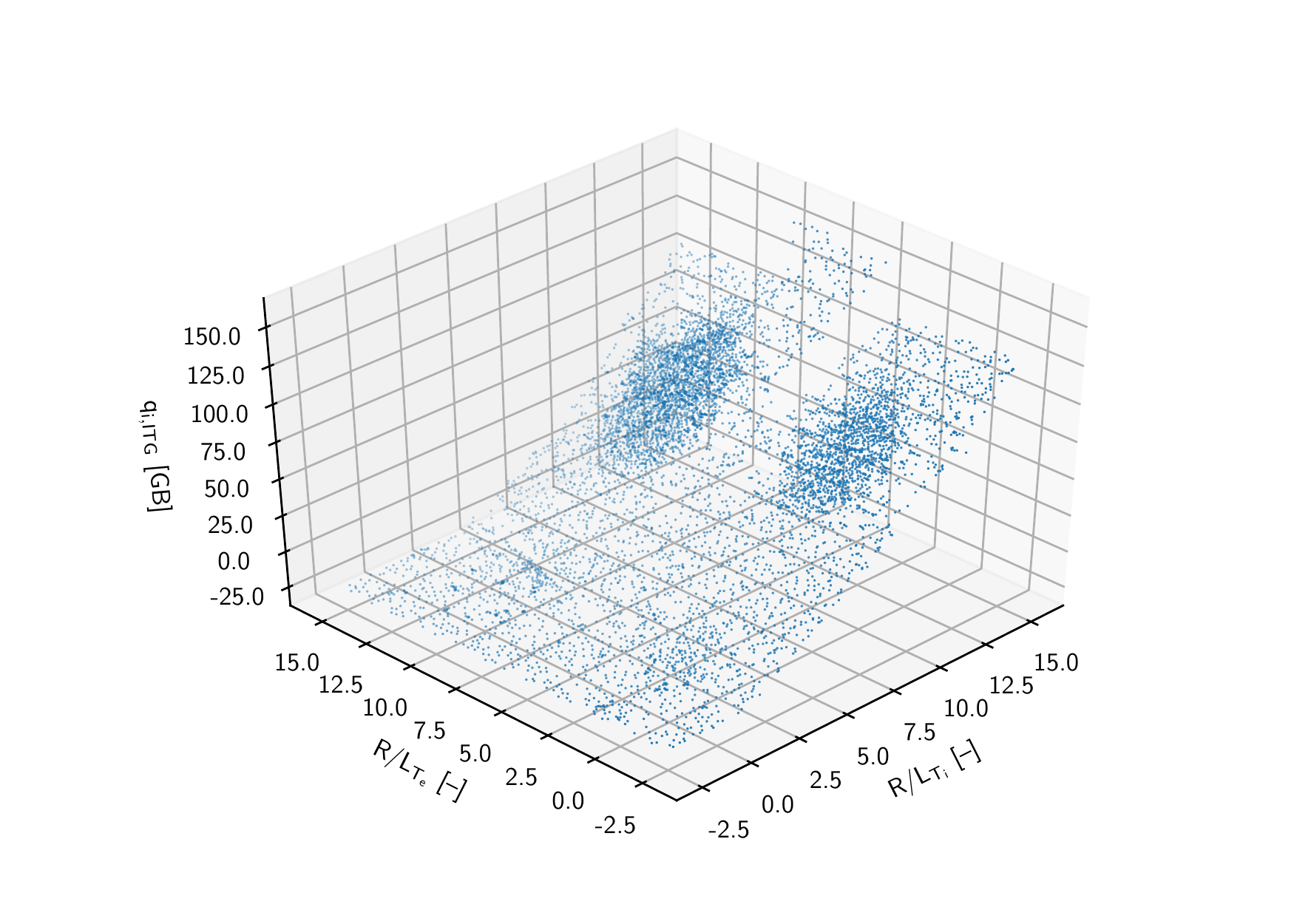}
	\caption{Reduced 3D sample dataset using $\zeta=1$, applying an overall reduction factor, $f=8.7$. As shown, the density distributions within the high-density regions have been reduced dramatically without significantly impacting the density in the lower-density background. The two high-density regions at low $R/L_{T_i}$ are almost visually indistinguishable from the background.}
	\label{fig:3DReductionDensity}
\end{figure}

An example of a reduced 3D dataset can be seen in Figures~\ref{fig:3DReductionUniform} and \ref{fig:3DReductionDensity}, using the same code inputs with the exception of $\zeta=0$ and $\zeta=1$, respectively. As shown in the figures, the overall reduction factor is larger for the density-dependent $k$ configuration ($f=10$) as opposed to the global $k$ configuration ($f=8$). For both sets, the core features and distributions of the dataset are still visible, showing that it is possible to reduce a dataset using the two-step clustering algorithm. 


\begin{figure}[tb]
	\centering
	\includegraphics[scale=0.7]{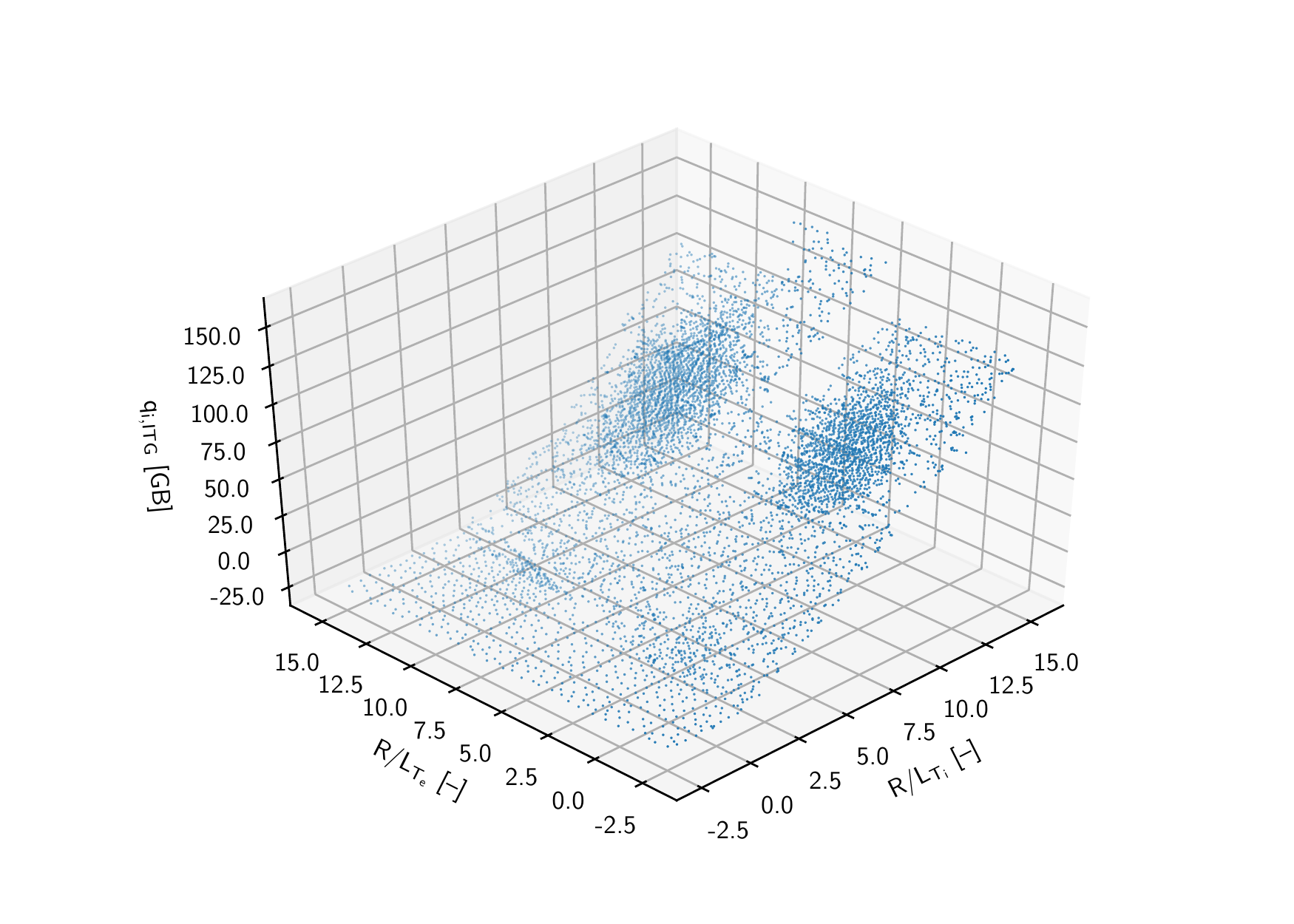}
	\caption{Reduced 3D sample dataset using $\zeta=1$ and $\epsilon_{min}=0.2$, applying an overall reduction factor, $f=10.0$. As shown, the density distributions within the two upper regions are more uniform due to the change in $\epsilon_{min}$, demonstrating its ability to remove high density spikes resulting from highly repetitive events occurring in the data.}
	\label{fig:3DReductionFlatten}
\end{figure}

However, given that the two high-density regions near $R/L_{Ti}=0.0$ in Figure~\ref{fig:3DReductionDensity} are nearly imperceptible, it is likely possible to use a combination of $\zeta$ and $\epsilon_{min}$ to achieve an output with a uniform density. For reasons of exposition, an example of this being targeted to flatten the density of these two regions is shown in Figure~\ref{fig:3DReductionFlatten}.

\subsubsection{3D dataset neural network training}
\label{subsubsec:Training3D}

The next step is the training NNs on these reduced datasets, using the same feed-forward NN model and training pipeline used previously~\cite{VanDePlassche2020}. For each dataset, 10 individual NNs are trained to form a committee of NNs. This allows an estimate of the regression uncertainties due to the random initialization of the training algorithm~\cite{Ho2021}. The NN contains 3 hidden layers, having 30 neurons per layer and each neuron using the $\tanh$ activation function. The Adam optimization algorithm and recommended settings were used in the training pipeline. This standardization of the training methodology within this study allows any noticeable trends in regression performance to be attributed primarily to the reduction of the training dataset.

The final step is to compare the NNs trained using the reduced dataset to the one trained using the original dataset. This is done using the two standard NN training metrics, namely the root-mean-squared error (RMSE) and the mean absolute error (MAE). These values are evaluated for each individual NN and then combined into a single mean and standard deviation per reduction factor, $f$, varied within this exercise by altering the global reduction ratio parameter, $a$. The remaining parameters were set to the recommended values listed in Table~\ref{tbl:RecommendedInputParameterValues} along with $MinPts=2$ and $n_{\text{min}}=6$. The difference between these values and the recommended values are because the latter have been refined since this test was performed, with the refined values shown in the table. Additionally, the RMSE and MAE metrics are computed with respect to the entirety of the original ``non-reduced" dataset, in order to allow the impact on regression quality to be separated from the overall statistics of the dataset when interpreting the results.

\begin{figure}[tb]
	\centering
	\includegraphics[width=0.97\columnwidth]{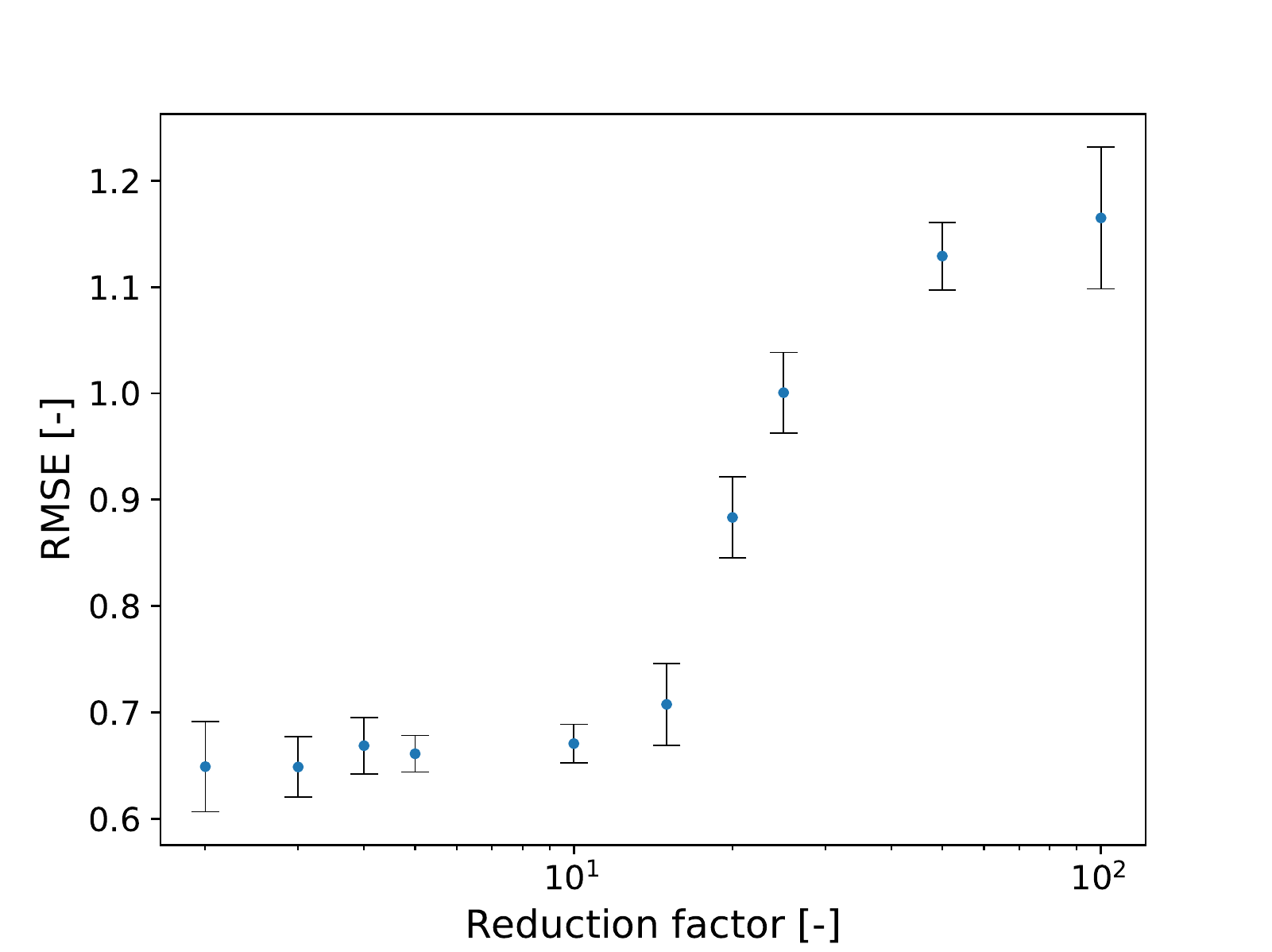}
	\caption{Root mean squared error (RMSE) for the surrogate models trained using the reduced 3D dataset compared against one trained using the non-reduced dataset, evaluated across all of the input points in non-reduced dataset. By interpreting this metric as a measure of the general reproducability of the surrogate model using the reduced dataset, a significant loss in model accuracy is observed at a specific overall reduction factor, $f\simeq15$.}
	\label{fig:RMSE3DReduced}
\end{figure}

\begin{figure}[tb]
	\centering
	\includegraphics[width=0.97\columnwidth]{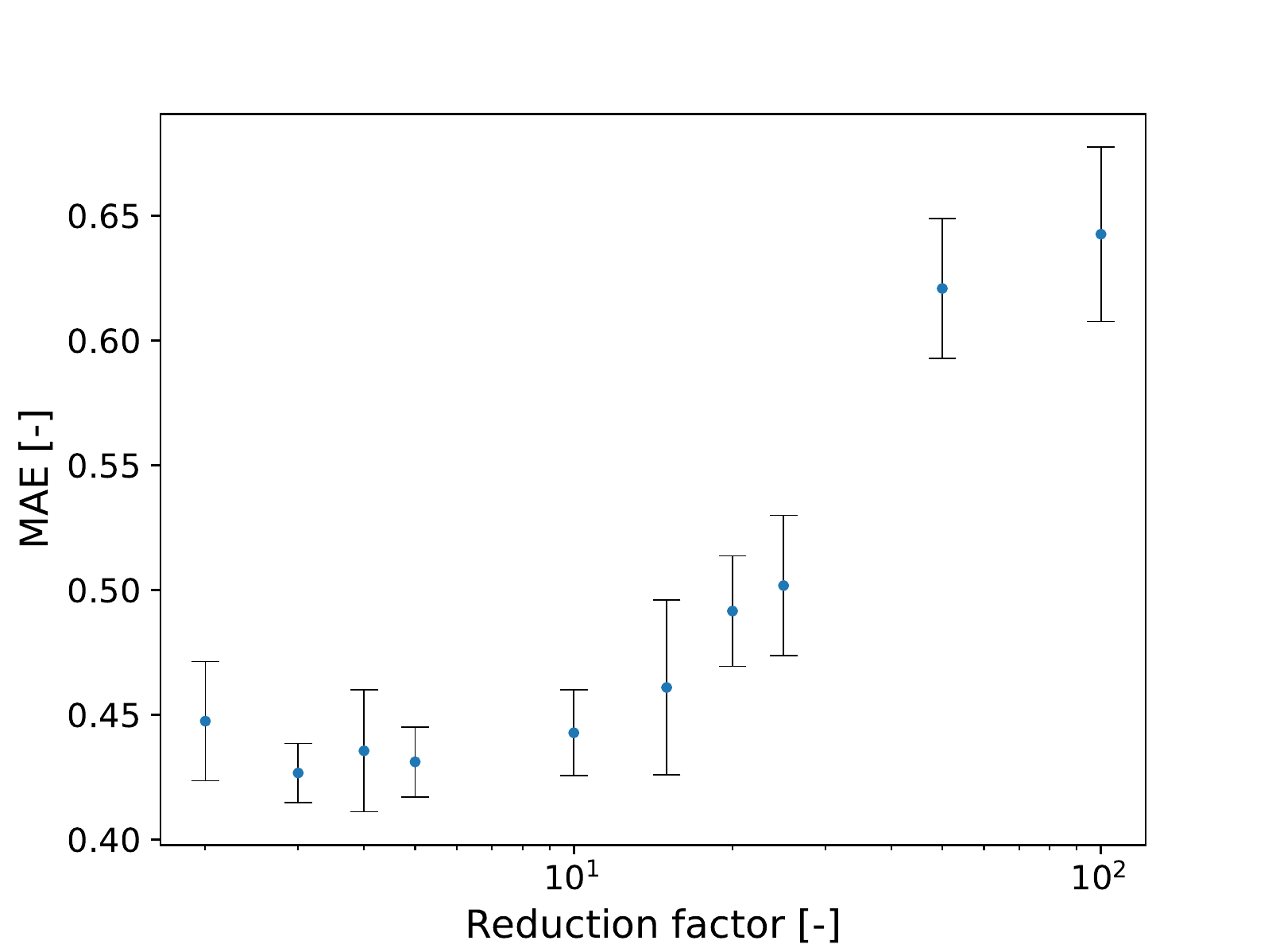}
	\caption{Mean absolute error (MAE) for the surrogate models trained using the reduced 3D dataset compared against one trained using the non-reduced 3D dataset, evaluated across all of the input points in the non-reduced dataset. By interpreting this metric as a measure of the general reproducability of the surrogate model using the reduced dataset, a significant loss in model accuracy is observed at a specific overall reduction factor, $f\simeq15$.}
	\label{fig:MAE3DReduced}
\end{figure}

Figures~\ref{fig:RMSE3DReduced} and \ref{fig:MAE3DReduced} display the RMSE and the MAE, respectively, as function of the reduction factor, $f$. For both metrics, no significant increase in error is detected at relatively low reduction factors. This is taken to imply that the reduced dataset retains the key features needed for an equivalent NN regression to be trained are still present. For this dataset, a steady increase in RMSE and MAE is detected for $f \ge 15$, showing that the two-step algorithm is reducing so much that essential dataset features for an equivalent NN regression are no longer captured. This result shows that the original dataset is likely over-populated for the features present in the underlying functions for the purposes of training NN regressions.


\subsection{Extension to target 16D application}
\label{subsec:TestCase16D}

The experimental data used in this test case was provided by the Joint European Torus (JET) experimental fusion plasma device, located in Culham, UK. This data is processed as described in Ref.~\cite{Ho2019}. This has led to a dataset with 15 input dimensions, which are then used to calculate particle, heat and momentum fluxes with the use of QuaLiKiz \cite{Bourdelle2007}. For the proof-of-principle extension, the ion heat flux, $q_{i,\text{ITG}}$, was again chosen as the single output dimension, making this a 16D dataset. The two-step-clustering algorithm was applied to this dataset to reduce it and additional NN surrogate models were trained from those reduced datasets.

\begin{figure}[tb]
	\centering
	\includegraphics[width=0.97\columnwidth]{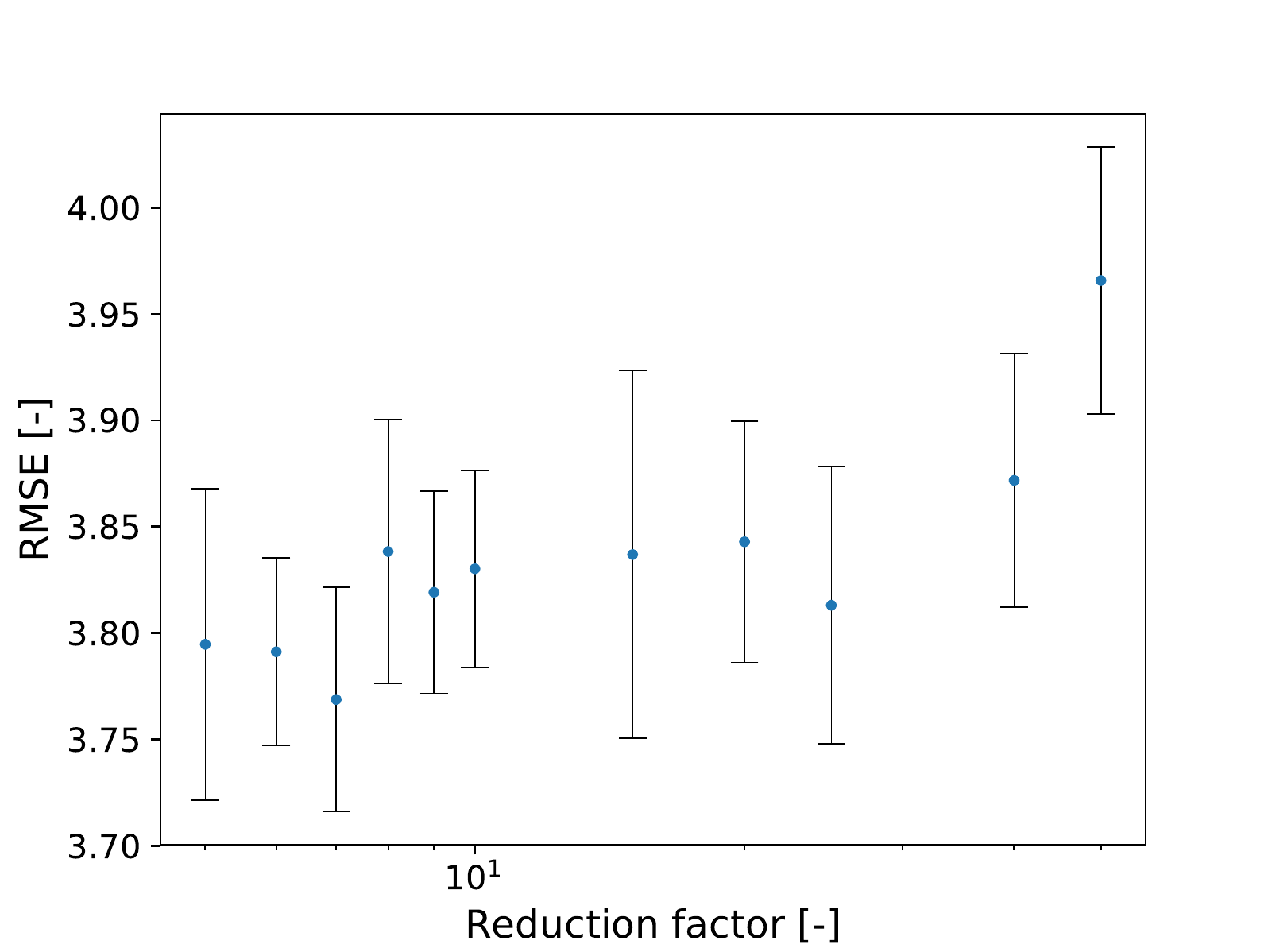}
	\caption{Root mean squared error (RMSE) for the surrogate models trained using the reduced 16D dataset compared against one trained using the 16D non-reduced dataset, evaluated across all of the input points in non-reduced dataset. By interpreting this metric as a measure of the general reproducability of the surrogate model using the reduced dataset, a significant loss in model accuracy is observed at a specific overall reduction factor, $f\simeq25$.}
	\label{fig:RMSE16DReduced}
\end{figure}

\begin{figure}[tb]
	\centering
	\includegraphics[width=0.97\columnwidth]{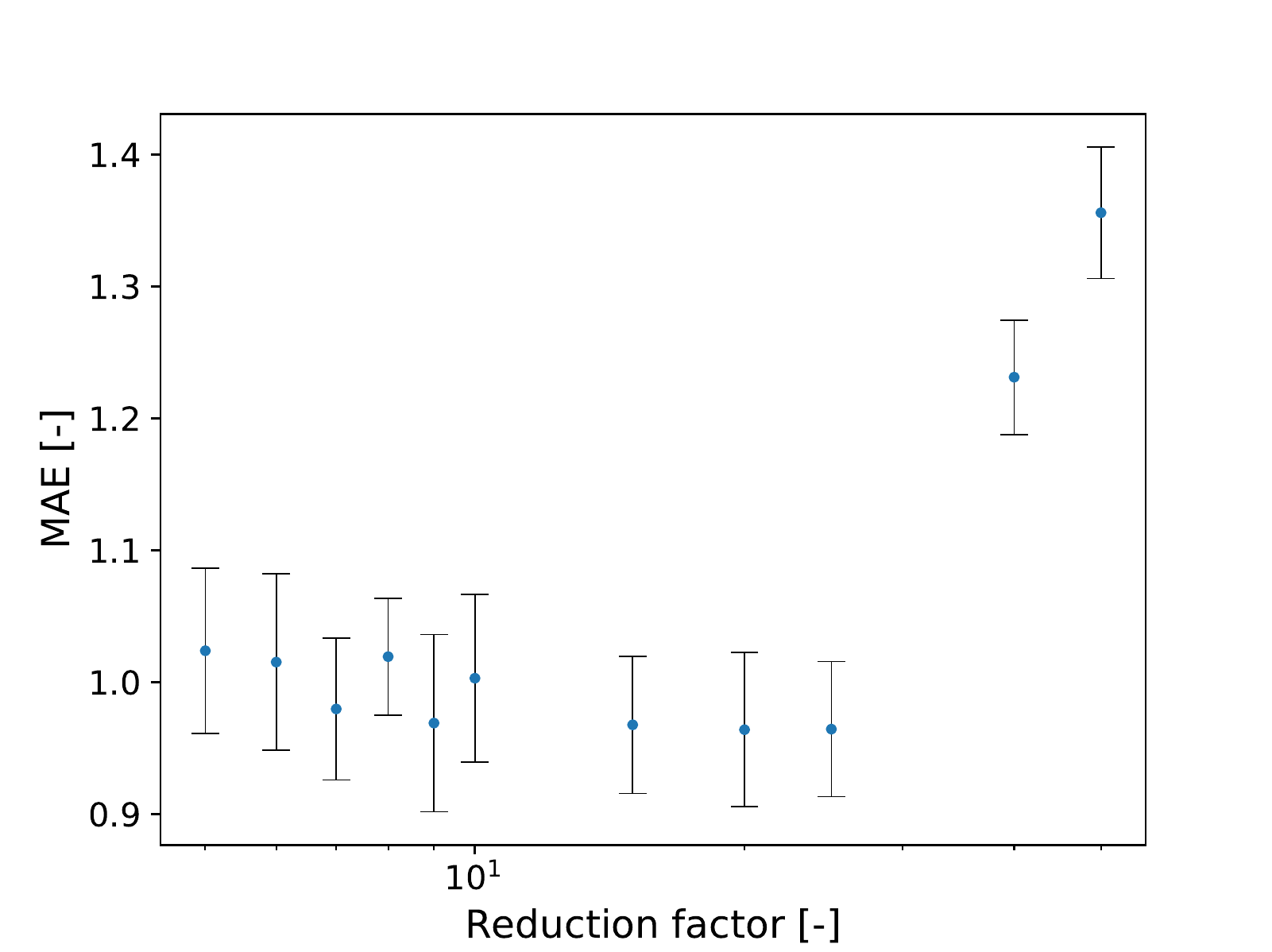}
	\caption{Mean absolute error (MAE) for the surrogate models trained using the reduced 16D dataset compared against one trained using the 16D non-reduced dataset, evaluated across all of the input points in non-reduced dataset. By interpreting this metric as a measure of the general reproducability of the surrogate model using the reduced dataset, a significant loss in model accuracy is observed at a specific overall reduction factor, $f\simeq25$.}
	\label{fig:MAE16DReduced}
\end{figure}

Similar to the previous 3D toy case, these ``clustered" NNs were then compared to the original model via the RMSE and MAE metrics computed with respect to the entire non-reduced dataset. Also similarly, the reduction algorithm was executed over a number of different reduction  factors by varying the global reduction ratio parameter, $a$, while setting the remaining parameters according to the recommendations in Table~\ref{tbl:RecommendedInputParameterValues} along with $MinPts=2$ and $n_{\text{min}}=6$. As shown in Figures~\ref{fig:RMSE16DReduced} and \ref{fig:MAE16DReduced}, the resulting RMSE and MAE metrics exhibit similar features except with increase occurring for $f \ge 25$.



\begin{figure}[tb]
	\centering
	\includegraphics[scale=0.7]{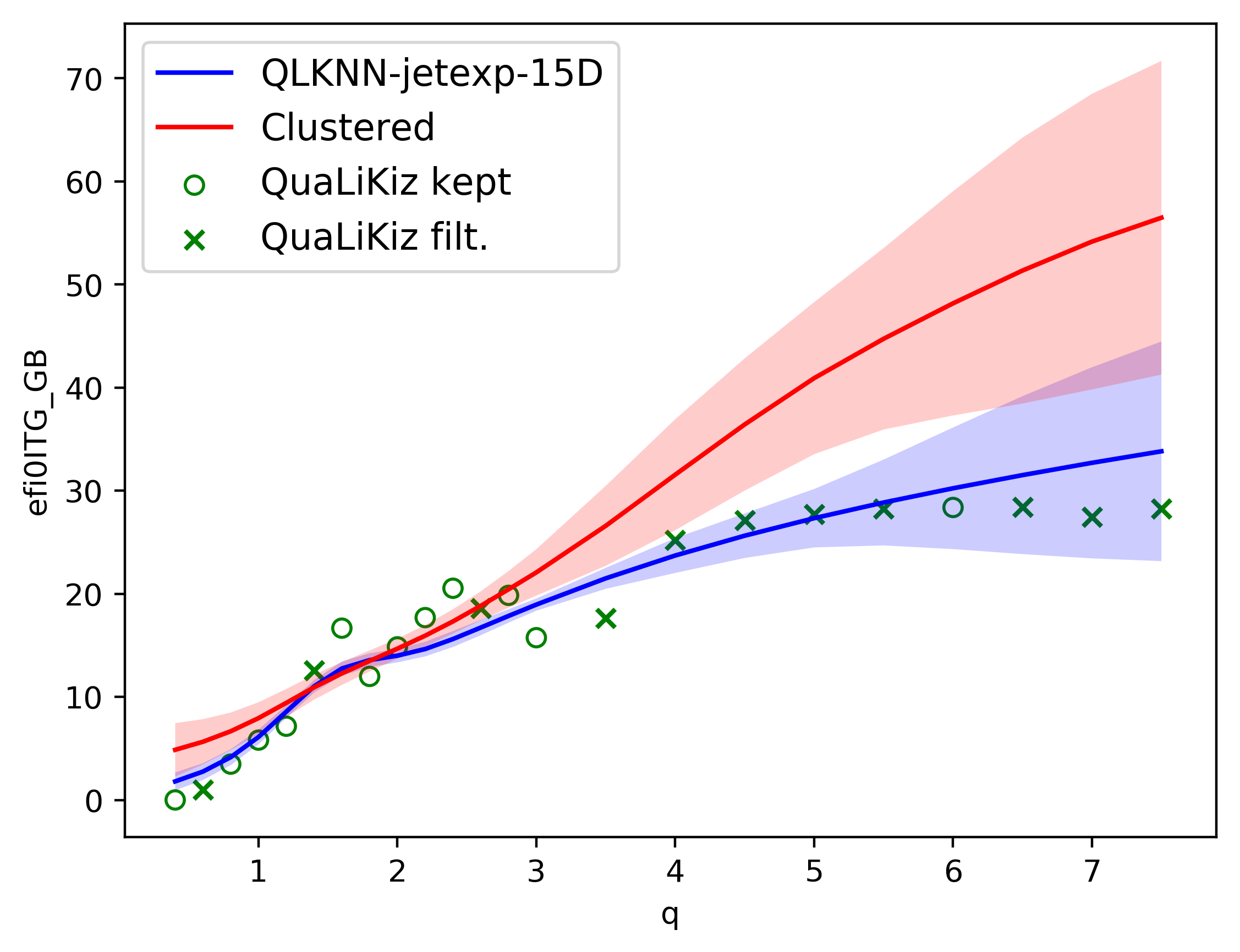}
	\caption{Comparison of the ion heat flux as a function of one input parameter in the 15D dataset, $q$, between the original NN (blue line), reduced NN (red line) and the underlying physics model (green points). The reduction did not significantly affect the high-density regions, indicated by the locations with small uncertainty ranges, but did reduce the fit quality in the low-density region. A reduction of $f\simeq25$ was applied to the data to produce this result.}
	\label{fig:QuaLiKizNNComparison}
\end{figure}

A more direct comparison of the two NNs at the elbow ($f\simeq25$) can be seen in Figure~\ref{fig:QuaLiKizNNComparison}. This application uses an ensemble of NNs, each identical in architecture but trained on the same dataset except with a different initialization. The training dataset was filtered in a way to ensure that high-density regions in the dataset are strongly correlated with low ensemble variance~\cite{Ho2021}. As expected, there is good agreement between the NNs trained on the original and the reduced dataset where the predicted variance is the lowest. However, the discrepancy between the two NN ensembles are most noticeable in the low-density regions, where the noise filtering in the reduction process likely removed a significant portion of data. While it can be argued that this noise filtering has improved the match between the NN and the original data in the high-density region, the overall appropriateness of this noise filtering is strongly dependent on the application and its investigation is outside the scope of this study. However, the ability to adjust $\epsilon_0$ when running the algorithm allows some control over the denoising process for such an investigation.

\section{Conclusions}
\label{sec:Conclusions}

In this work, a two-step algorithm is proposed for the detection and reduction of high data density regions within a larger $N$-dimensional dataset. This algorithm is based on a combination of the widely-used clustering algorithms, DBSCAN and $k$-means, applied sequentially and iteratively to determine an density-dependent reduction ratio. The DBSCAN algorithm is first used to detect clusters of high data density, and the information about those clusters is used by $k$-means to partition them into representative regions for reduction. This work is novel not only in that it proposes a robust data detection and reduction scheme, but also proposes using DBSCAN iteratively to distinguish between regions of varying density.

An application of the two-step clustering reduction algorithm was performed, reducing the training dataset previous used for training NN surrogate models of fusion plasma turbulent transport phenomena. The resulting analysis showed that the dataset in this application could potentially be reduced by a factor of $\sim$20 before significantly impacting the accuracy of the NN surrogate model. This has implications in lowering the computation expense of generating these datasets and/or expanding the parameter space included in such dataset. The method is also envisioned to help with maintaining a living dataset, in which new data can be continuously accumulated and trimmed without fear of unintentionally biasing the dataset. Future work can be performed to improve the generalizability and robustness of the $k$ calculation, in order to remove extraneous parameters to the algorithm. One such example is implementing an internal calculation of $\tau$ such that $\zeta$ moves between applying an approximately uniform reduction factor to enforcing an approximately uniform dataset density. Overall, the authors see many potential use-cases for this algorithm in various machine learning applications where the dataset is either collected on-the-fly or via a top-down approach.




\section{Source Code}
The complete code can be found at: 
\url{\GitHubLoc} and
\url{\GitHubLocSecond}.

\printbibliography

\end{document}